\documentclass[10pt]{revtex4}
\usepackage{amsthm}
\usepackage{amssymb}
\usepackage{amsmath}
\usepackage{graphicx}
\def \D {\tilde{\nabla}}
\def\l{\label}

\def\Th{\Theta}

\def\sig{\sigma}
\def\om{\omega}
\def\udot{\dot{u}}
\def\nab{\nabla}
\def\3nab{\tilde{\nabla}}

\def\lgl{\langle}
\def\rgl{\rangle}
\def\la {\langle}
\def\ra {\rangle}
\def\nn{\nonumber}
\def\c{\mbox{curl}}

\def\hsp5{\hspace{5mm}}
\newcommand{\sfrac}[2]{{\textstyle{#1\over#2}}}
\def\case#1/#2{\textstyle\frac{#1}{#2}}

\def\be {\begin{equation}}
\def\ee {\end{equation}}
\def\ber {\begin{eqnarray}}
\def\eer {\end{eqnarray}}
\def\bea {\begin{eqnarray}}
\def\eea {\end{eqnarray}}

\def\bc {\begin{center}}
\def\ec {\end{center}}
\newcommand{\hs}{\,-\,}

\def\cqg{{\it Class. Quantum Grav.}\ }

\def\grg{{\it Gen. Relativ. Grav.}\ }

\def\etal\;{{\it et al.}}

\newcommand{\0}{^{(0)}}
\newcommand{\1}{^{(1)}}
\newcommand{\2}{^{(2)}}

\begin{document}

\title{The evolution of cosmological gravitational waves in $f(R)$ gravity}

\author{K. N. Ananda\footnote{\tt
kishore.ananda@gmail.com}}\affiliation{Department of Mathematics and
Applied Mathematics, University of Cape Town, South Africa.}
\author{S. Carloni\footnote{\tt
sante.carloni@gmail.com}}\affiliation{Department of Mathematics and
Applied Mathematics, University of Cape Town, South Africa.}
\author{P. K. S. Dunsby\footnote{\tt peter.dunsby@uct.ac.za}}
\affiliation{Department of Mathematics and Applied Mathematics,
University of Cape Town, South Africa.} \affiliation{South African
Astronomical Observatory, Observatory Cape Town, South Africa.}

\begin{abstract}
We give a rigorous and mathematically clear presentation of the
Covariant and Gauge Invariant theory of gravitational waves in a
perturbed Friedmann-Lema\^{\i}tre-Robertson-Walker universe for
Fourth Order Gravity, where the matter is described by a perfect
fluid with a barotropic equation of state.
As an example of a consistent analysis of tensor perturbations
in Fourth Order Gravity, we apply the formalism to a simple
background solution of $R^n$ gravity.
%
We obtain the exact solutions of the perturbation equations for
scales much bigger than and smaller than the Hubble radius. It is
shown that the evolution of tensor modes is highly sensitive to the
choice of $n$ and an interesting new feature arises. During the 
radiation dominated era, their exist a growing tensor perturbation 
for nearly all choices of $n$. This occurs even when the background 
model is undergoing accelerated expansion as opposed to the case 
of General Relativity. Consequently, cosmological gravitational wave 
modes can in principle provide a strong constraint on the theory of gravity
independent of other cosmological data sets.
\end{abstract}

\date{\today}
 \maketitle

\section{Introduction}

In the near future, Gravitational Waves (GW) will become a very
important source of data in cosmology. Cosmological GW are produced
at very early times in the evolution of the universe and almost
immediately decouple from the cosmic fluid. Consequently they carry
information about the conditions that existed at this time, thus
providing a way of constraining models of inflation
\cite{Grishchuk:1974ny}.

Even if GW are decoupled from the cosmic fluid their presence still
influences some features of the observable universe. In particular,
a GW background will produce a signature that can be found in the
anisotropies  \cite{Rubakov:1982df}  and polarization
\cite{Hu:1997hv} of the Cosmic Microwaves Background (CMB). This,
together with the remarkable improvements in the sensitivity of CMB
measurements, opens the possibility of obtaining  important
information about GW in an indirect way.

In the last few years the idea of a geometrical origin for Dark
Energy (DE) i.e. the connection between DE and a non-standard
behavior of gravitation on cosmological scales has attracted a
considerable amount of interest.

Higher order gravity, and in particular fourth order gravity, has
been widely studied in the case of the
Friedmann-Lema\^{\i}tre-Robertson-Walker (FLRW) metric using a
number of different techniques (see for example \cite{revnostra,
Odintsov, Carroll, Capozziello:2005ku, Capozziello:2006dj,
Capozziello:2006ph, Capozziello:2006jj}). Recently a general
approach was developed to analyze the phase space of the fourth
order cosmologies \cite{cdct:dynsys05,SanteGenDynSys,shosho},
providing for the first time a way of obtaining exact solutions
together with their stability and a general idea of the qualitative
behavior of these cosmological models.

The phase space analysis shows that for FLRW models there exist
classes of fourth order theories in which the cosmology evolves
naturally towards an accelerated expansion phase which can be
associated with a DE-like era. Although this feature is particularly
attractive, a problem connected with the use of these theories is
that there is too much freedom in the form of the theory itself.
Consequently, it is crucial to investigate these models in some
detail in order to devise observational constraints  which are able
to eliminate this degeneracy.

A key step in this process
is the development of a full theory of cosmological perturbations. A
detailed analysis of the evolution of the scalar perturbations on
large scales has recently been given in \cite{Carloni:2007yv}. Here
we will focus on the evolution of the tensor perturbations, which are
related to GWs.
This is motivated by the well known fact \cite{Will}, that the
features of GWs in GR are rather special and therefore the detection
of any deviation from this behavior would be a genuine proof of the
break down of standard GR.

The aim of this paper is to present a general framework within
which to consistently analyze tensor perturbations of FLRW models in
fourth order gravity (see \cite{current1,current2,
current3,current4} for other recent contributions to this area). As
an explicit example we apply our approach to the case of $R^n$
gravity. We investigate the possible constraints one can place on
such a model through future observations of gravitational waves
independently of existing cosmological data.


In order to achieve this goal a perturbation formalism needs to be
chosen that is best suited for this task. One possible choice is the
Bardeen metric based approach \cite{bardeen, KodamaSasaki,
Bean2007} which guarantees the gauge invariance of the results.
However this approach has the drawback of introducing variables
which only have a clear physical meaning in certain gauges
\cite{BDE}.  Although this is not a big problem in the context of
General Relativity (GR), this is not necessarily true in the case of
higher order gravity and consequently can lead to a
miss-interpretation of the results.

In what follows we will use, instead, the covariant and gauge
invariant approach developed for GR in \cite{EB, EBH, BDE, DBE, BED,
DBBE} which has the advantage of using perturbation variables with a
clear geometrical and physical interpretation. We take advantage of
the fact that in this approach the non-Einstein part of the
gravitational interaction can be considered as an effective fluid (the
{\it curvature fluid}) coupled with standard matter. This specific
recasting of the field equations makes the development of
cosmological perturbation theory even more transparent.

The main results of the paper are as follows. (1) We find that the
evolution of tensor modes  is extremely sensitive to the choice of
$f(R)$ theory. (2) In the specific case of  $R^n$ gravity, the
tensor modes are in general weaker due to a higher expansion rate in
the background. (3) During the radiation dominated era, their exist 
a growing tensor mode for nearly all interesting values of $n$.

The paper is organized as follows. In section II we give a brief
review of the 1+3 gauge invariant covariant approach in a general
setting. In section III we present the equations necessary for the
study of linear tensor perturbations for a general imperfect fluid.
In section IV we investigate how these equations are modified when
considering fourth order gravity. In section V we adapt these
equations for the specific case of $R^n$ gravity and study tensor
perturbations both in vacuum and in the presences of dust/radiation
fluid. Finally, we present our discussions and conclusions in
section VI.

\section{The 1+3 covariant approach to cosmology}

The starting point (and the corner stone) of our analysis is the 1+3
covariant approach to cosmology \cite{EllisCovariant}. This approach
consists of deriving  a set of first order differential equations
and constraints for some suitable, geometrically well defined
quantities (the {\em 1+3 equations}) that are completely equivalent
to the Einstein field equations. This has the advantage of
simplifying the analysis of general spacetimes which can be foliated
as a set of three dimensional (spacelike) surfaces. In the following
we give a very brief introduction to the parts of this formalism
used in this paper.

\subsection{Preliminaries}

We will adopt natural units ($\hbar=c=k_{B}=8\pi G=1$) throughout
this paper, Latin indices run from 0 to 3. The symbol $\nabla$
represents the usual covariant derivative and $\partial$ corresponds
to partial differentiation. We use the $-,+,+,+$ signature and the
Riemann tensor is defined by
\begin{equation}
R^{a}{}_{bcd}=W^a{}_{bd,c}-W^a{}_{bc,d}+ W^e{}_{bd}W^a{}_{ce}-
W^f{}_{bc}W^a{}_{df}\;,
\end{equation}
where the $W^a{}_{bd}$ is the usual Christoffel symbol (i.e.
symmetric in the lower indices), defined by
\begin{equation}
W^a{}_{bd}=\sfrac{1}{2}g^{ae}
\left(g_{be,d}+g_{ed,b}-g_{bd,e}\right)\;.
\end{equation}
The Ricci tensor is obtained by contracting the {\em first} and the
{\em third} indices
\begin{equation}\label{Ricci}
R_{ab}=g^{cd}R_{cadb}\;.
\end{equation}
Finally the Einstein-Hilbert action in the presence of matter is
defined as
\begin{eqnarray}
\mathcal{A}=\int d x^{4}\sqrt{-g}\left[\sfrac{1}{2}R+\mathcal{L}_m
\right]\;.
\end{eqnarray}

\subsection{Kinematics}

In order to derive the 1+3 equations we have to choose a set of
observers i.e. a 4-velocity field $u^a$. This choice depends
strictly on the theory of gravity that we are treating. In this
section we give the set of equations for a general velocity field.
In later sections we will discuss how this situation is modified in
the case of f(R) gravity.

Given the velocity $u^a$, we can define the projection tensor into
the tangent 3-spaces orthogonal to the flow vector:
\begin{equation}
h_{ab}  \equiv g_{ab}+u_au_b\; \Rightarrow h^a{}_b
h^b{}_c=h^a{}_c\;, ~h_{ab}u^b=0,
\end{equation}
and the kinematical quantities can be obtained by splitting the
covariant derivative of $u_a$ into its irreducible parts:
\begin{equation}
\nab_b u_a=\3nab_b u_a-A_a u_b\;, ~~~\3nab_b u_a=\sfrac{1}{3}\Theta
h_{ab} +\sigma_{ab}+\omega_{ab}\;, \label{eq:dec}
\end{equation}
where $\3nab_a$ is the spatially totally projected covariant
derivative operator orthogonal to $u^a$, $A_a = \dot{u}_a$ is the
acceleration ($A_b u^b=0$), $\Theta$ is the expansion parameter,
$\sigma_{ab}$ the shear ($\sigma_{ab} =\sigma_{(ab)}$, $\sigma^a{}_a
= \sigma_{ab}u^b=0$) and $\omega_{ab}$ is the vorticity
($\omega_{ab} =\omega_{[ab]}$, $\omega_{ab}u^b=0)$. Following the
standard convention we will indicate the symmetrization over two
indices of a tensor with round brackets and the antisymmetrization
with square ones.

In the $u^a$ frame, the {\em Weyl or conformal curvature tensor}
$C_{abcd}$ can be split into its Electric ($E_{ab}$) and Magnetic
($H_{ab}$) components respectively:
\bea E_{ab} = C_{acbd}\,u^c\,u^d & \Rightarrow &
E^a{}_a = 0 \ , ~E_{ab} = E_{(ab)} \ , ~E_{ab}\,u^b = 0 \ , \\
H_{ab} = {\sfrac12}\,\eta_{ade}\,C^{de}{}_{bc}\,u^{c} & \Rightarrow
& H^a{}_a = 0 \ , ~H_{ab} = H_{(ab)} \ , ~H_{ab}\,u^b = 0 \ . \eea
In what follows we will use orthogonal projections of vectors and
orthogonally projected symmetric trace-free part of tensors. They
are defined as follows:
\be v^{\la a\ra} = h^{a}{}_{b}\,v^{b} \ , \hsp5 X^{\la ab\ra} = [\
h^{(a}{}_{c}\,h^{b)}{}_{d} - \sfrac{1}{3}\,h^{ab}\,h_{cd}\ ]\,X^{cd}
\ . \ee
Angle brackets may also be used to denote orthogonal
projections of covariant time derivatives along $u^{a}$:
\be \dot{v}{}^{\la a\ra} = h^{a}{}_{b}\,\dot{v}{}^{b} \ , \hsp5
\dot{X}{}^{\la ab\ra} = [\ h^{(a}{}_{c}\,h^{b)}{}_{d} -
\sfrac{1}{3}\,h^{ab}\,h_{cd}\ ]\,\dot{X}{}^{cd} \ .
\ee
\subsection{Energy-momentum tensors}

The choice of frame, i.e., choice of velocity field $u^{a}$ and
therefore the projection tensor $h_{ab}$ allows one to obtain an
irreducible decomposition of a generic Energy-Momentum Tensor (EMT),
$T_{a b}^{tot}$. The following unbarred quantities have been
derived from the total EMT, quantities relating to the effective
fluids will be denoted with sub/super-scripts in order help to avoid
confusion in later sections and to generalize to a multi-fluid system.
\begin{equation}
T_{ab}^{tot}={\mu} u_a u_b +{p} h_{ab}+{q}_a
u_b+{q}_b u_a+{\pi}_{ab}\;, \label{eq:pf}
\end{equation}
where ${\mu}$ is the total energy density and ${p}$ is the
total isotropic pressure of the fluid, ${q}_a$ represents the
total energy flux, ${\pi}_{ab}$ is the total anisotropic
pressure. Additionally, we have the following constraints
\bea \hsp5 {q}_a\,u^a = 0 \ , ~{\pi}^a{}_a = 0 \ ,
~{\pi}_{ab} = {\pi}_{(ab)} \ , ~{\pi}_{ab}\,u^b = 0\;.
\nonumber \eea

The various components of the total energy momentum tensor can be
isolated in the following way:
\begin{eqnarray}\label{totEMTcomp}
{\mu} &=& T_{ab}^{tot} u^{a} u^{b}, \\
\nonumber\\
{p} &=& \frac{1}{3} T_{ab}^{tot} h^{ab}, \\
\nonumber\\
{q}_{a} &=& -T_{cd}^{tot} u^{c} h^{d}{}_{a}, \\
\nonumber\\
{\pi}_{ab} &=& T_{\langle ab \rangle}^{tot}.
\end{eqnarray}

In a general fluid the pressure, energy density and  entropy are
related to each other by an equation of state $p=p(\mu,s)$. A fluid
is considered {\em perfect} if $q_a$ and $\pi_{ab}$ vanish, and {\em
barotropic} if the entropy is a constant i.e. the equation of state
reduces to $p=p(\mu)$.

\subsection{Propagation and constraint equations}

Writing the Ricci and the Bianchi identities in terms of the $1+3$
variables defined above, we obtain a set of evolution equations
(here the `curl' is defined as $(\c\,X)^{ab} = \eta^{cd\lgl
a}\,\3nab_{c}X^{b\rgl}\!_{d}$) :
\be \l{eq:ray} \dot{\Th} - \3nab_{a}\udot^{a} = -
\,\sfrac{1}{3}\,\Th^{2} + (\udot_{a}\udot^{a}) - 2\,\sig^{2} +
2\,\om^{2} - \sfrac{1}{2}\,({\mu}+3{p}) \ , \ee
\be \l{eq:omdot} \dot{\om}^{\lgl a\rgl} - \sfrac{1}{2}\,\eta^{abc}\,
\3nab_{b}\udot_{c} = - \,\sfrac{2}{3}\,\Th\,\om^{a} +
\sig^{a}\!_{b}\,\om^{b}  ,\ee
\bea \l{eq:sigdot} \dot{\sig}^{\lgl ab\rgl} - \3nab{}^{\lgl
a}\udot^{b\rgl} = - \,\sfrac{2}{3}\,\Th\,\sig^{ab} + \udot^{\lgl
a}\, \udot^{b\rgl} - \sig^{\lgl a}\!_{c}\,\sig^{b\rgl c} - \om^{\lgl
a}\,\om^{b\rgl} - (E^{ab}-\sfrac{1}{2}\,{\pi}^{ab}) \ , \eea
\bea (\dot{E}^{\lgl ab\rgl}+\sfrac{1}{2}\,\dot{{\pi}}^{\lgl
ab\rgl}) - (\c\,H)^{ab} + \sfrac{1}{2}\,\3nab^{\lgl
a}{q}^{b\rgl} & = & -
\,\sfrac{1}{2}\,({\mu}+{p})\,\sig^{ab}
- \Th\,(E^{ab}+\sfrac{1}{6}\,{\pi}^{ab}) \\
& &  + \ 3\,\sig^{\lgl a}\!_{c}\,(E^{b\rgl c}
-\sfrac{1}{6}\,{\pi}^{b\rgl c}) - \udot^{\lgl
a}\,{q}^{b\rgl}
\nonumber \\
& &  + \ \eta^{cd\lgl a}\,[\ 2\,\udot_{c}\,H^{b\rgl}\!_{d} +
\om_{c}\,(E^{b\rgl}\!_{d}+\sfrac{1}{2}\,{\pi}^{b\rgl}\!_{d})\ ]
\ , \nonumber \eea
\bea \dot{H}^{\lgl ab\rgl} + (\c\,E)^{ab}
-\sfrac12(\c\,{\pi})^{ab} & = & - \,\Th\,H^{ab} + 3\,\sig^{\lgl
a}\!_{c}\,H^{b\rgl c}
+ \sfrac{3}{2}\, \om^{\lgl a}\,{q}^{b\rgl} \\
& & \hsp5 - \ \eta^{cd\lgl a}\,[\ 2\,\udot_{c}\,E^{b\rgl}\!_{d} -
\sfrac{1}{2}\,\sig^{b\rgl}\!_{c}\,{q}_{d} -
\om_{c}\,H^{b\rgl}\!_{d}\ ] \ , \nonumber \eea
\be \l{eq:cons1} \dot{{\mu}} + \3nab_{a}{q}^{a} = -
\,\Th\,({\mu}+{p}) - 2\,(\udot_{a}{q}^{a}) -
(\sig^{a}\!_{b}{\pi}^{b}\!_{a})\;, \ee
\be \l{eq:cons2} \dot{{q}}^{\lgl a\rgl} + \3nab^{a}{p} +
\3nab_{b}{\pi}^{ab} = - \,\sfrac{4}{3}\,\Th\,{q}^{a} -
\sig^{a}\!_{b}\,{q}^{b} - ({\mu}+{p})\,\udot^{a} -
\udot_{b}\,{\pi}^{ab} - \eta^{abc}\,\om_{b}\,{q}_{c} \ , \ee
and a set of constraints
\bea \l{eq:divE}  \3nab_{b}(E^{ab}+\sfrac{1}{2}\,{\pi}^{ab}) -
\sfrac{1}{3}\,\3nab^{a}{\mu} + \sfrac{1}{3}\,\Th\,{q}^{a} -
\sfrac{1}{2}\,\sig^{a}\!_{b}\,{q}^{b} - 3\,\om_{b}\,H^{ab} - \
\eta^{abc}\,[\ \sig_{bd}\,H^{d}\!_{c} -
\sfrac{3}{2}\,\om_{b}\,{q}_{c}\ ]  = 0\;, \eea
\bea \l{eq:divH}\3nab_{b}H^{ab} + ({\mu}+{p})\,\om^{a} +
3\,\om_{b}\,(E^{ab}-\sfrac{1}{6}\,{\pi}^{ab})+ \ \eta^{abc}\,[\
\sfrac{1}{2}\,\3nab_{b}{q}_{c} + \sig_{bd}\,(E^{d}\!_{c}
+\sfrac{1}{2}\,{\pi}^{d}\!_{c})\ ] &=& 0\;, \eea
\be\l{eq:onu} \3nab_{b}\sig^{ab} - \sfrac{2}{3}\,\3nab^{a}\Th +
\eta^{abc}\,[\ \3nab_{b}\om_{c} + 2\,\udot_{b}\,\om_{c}\ ] +
{q}^{a} =0 \;, \ee
\be   \3nab_{a}\om^{a} - (\udot_{a}\om^{a}) = 0\;, \ee
\be \l{hconstr}  H^{ab} + 2\,\udot^{\lgl a}\, \om^{b\rgl} +
\3nab^{\lgl a}\om^{b\rgl} - (\c\,\sig)^{ab} = 0\;, \ee
that are completely equivalent to the Einstein equations. It is from
these equations that we derive the general evolution equations for
linear tensor perturbations.

\section{Tensor Perturbation Equations}
\subsection{The Background}

The equations presented in the previous section hold in any
spacetime we may wish to analyze. However, in what follows we will
focus on the class of spacetimes that can be thought of as
describing an "almost" Friedmann-Lema\^{i}tre-Robertson-Walker
(FLRW) model, motivated by the fact that current observations
suggest that the universe appears to deviate only slightly from
homogeneity and isotropy. We can define a FLRW spacetime in terms of
the variables above. Homogeneity and isotropy imply:
 \begin{equation}
 \sigma =\omega= 0\;, ~~\3nab_a f=0\;,
 \label{eq:rwcond1}
 \end{equation}
 where $f$ is any scalar quantity; in particular
 \begin{equation}
 \3nab_a{\mu} =\3nab_a {p}=0 ~~\Rightarrow~~, ~\dot{u}_a = 0  \;.
 \label{eq:rwcond3}
 \end{equation}
It follows that the governing equations for this background are
 \begin{equation}
\dot{\Theta}+\sfrac{1}{3}
\Theta^2+\sfrac{1}{2}\left({\mu}+3{p}\right)= 0\;,
\label{eq:rayback}
\end{equation}
\begin{equation}
\tilde{R} = 2\left[- \sfrac{1}{3} \Theta^2+{\mu}\right]\;,
\end{equation}
\begin{equation} \label{cons2}
    \dot{{\mu}}+\Theta\left({\mu}+{p}\right)=0\;.
\end{equation}
Now in order to describe small deviations from a FLRW spacetime we
simply take all the quantities that are zero in the background as
being first order, and retain in the equations
(Eq.~(\ref{eq:ray})-(\ref{hconstr})) only the terms that are linear
in these quantities, i.e. we drop all second order terms. This
procedure corresponds to the {\em linearization} in the 1+3
covariant approach and it greatly simplifies the system of
equations. In particular, the scalar, vector and tensor parts of the
perturbations are decoupled, so that we are able to treat them
separately. In what follows we will focus only on the tensor
perturbations.

\subsection{The general linear tensor perturbation equations}%

The 1+3 covariant description of gravitational waves in the context
of cosmology has been considered by~\cite{DBBE}. The linearized
gravitational waves are described by the transverse and trace-free
degrees of freedom once scalars have been switched off. Therefore,
focusing only on tensor perturbations the necessary evolution
equations are
\begin{equation}\label{eqsigma}
    \dot{\sigma} _{ab} + \frac{2}{3}\,\Theta \,\sigma _{ab}
  + E_{ab} - \frac{1}{2} {\pi}_{ab}=0\;,
\end{equation}
\begin{equation}\label{eqmag}
   \dot{H}_{ab}+
   H_{ab}\,\Theta+(\c\,E)_{ab} - \frac{1}{2}(\c\,{\pi})_{ab} = 0\;,
\end{equation}
\begin{equation}\label{eqelect}
   \dot{E}_{ab}+E_{ab}\,\Theta -
  (\c\,H)_{ab}  +
   \frac{1}{2}\left({\mu} +{p} \right) \,\sigma_{ab}
   + \frac{1}{6}\Theta\,{\pi}_{ab} +
   \frac{1}{2} \dot{{\pi}}_{ab}=0\;,
\end{equation}
together with the conditions
\begin{equation}\label{constraints}
\3nab_{b}H^{ab}=0\;,\quad \3nab_{b}E^{ab}=0\;,\quad
H_{ab}=(\c\,\sigma)_{ab}\;.
\end{equation}
Note that, since the linear tensor perturbations are
frame-invariant, the structure of the equations does not depend on
the choice of 4-velocity ,$u_{a}$. In the following, however, we
shall choose the frame associated with standard matter ($u_a=
u^{m}_a$). The motivation for such a choice is the fact that real
observers are attached to galaxies and these galaxies follow the
standard matter geodesics. Taking the time derivative of the above
equations we obtain
\begin{equation}\label{eq2ordSigma}
\ddot{\sigma}_{ab}-\3nab^{2}\sigma  + \frac{5}{3}\, \Theta \,
\dot{\sigma}_{ ab} +  \left(\frac{1}{9}\, {\Theta }^2 +
\frac{1}{6}{\mu} - \frac{3}{2}{p}  \right)\, \sigma_
{ab}=\dot{{\pi}}_{ab}+ \frac{2}{3}\, \Theta \, {\pi}_{ab}\;,
\end{equation}
\begin{equation}\label{eq2ordMag}
\ddot{H}_{ab} - \3nab^{2}H_{ab} + \frac{7}{3}\, \Theta\,
\dot{H}_{ab} + \frac{2}{3}\, \left({\Theta }^2 - 3 {p} \right)\,
H_{ab} = (\c\, \dot{{\pi}})_{ab}+\frac{2 }{3}\, \Theta\, (\c\, \
{\pi})_{ab}\;,
\end{equation}
\begin{eqnarray}\label{eq2ordElect}
\nn \ddot{E}_{ab}&-&\3nab^{2}E_{ab}+
\frac{7}{3}\,\Theta\,\dot{E}_{ab} +\frac{2}{3}\, \left( {\Theta}^2 -
3 {p} \right) {E}_{ab}  -  \frac{1 }{6}\Theta\,\left({\mu}
+ {p} \right) \,\left( 1 + 3\,c_{s}^{2}\right)\,\sigma _{ab} \\
&=& -
\left[\frac{1}{2}\ddot{{\pi}}_{ab}-\frac{1}{2}\3nab^{2}{\pi}_{ab}
+\frac{5}{6}\,\Theta\,\dot{{\pi}}_{ab} + \frac{1}{3}\left(
{\Theta }^2 - {\mu} \right) \,{\pi}_{ab}\right]\;,
\end{eqnarray}
where $c_{s}^{2}=\dot{p}/\dot{\mu}$ and we have used the
Raychaudhuri equation (Eq.~(\ref{eq:ray})), the energy conservation
equation (Eq.~(\ref{cons2})) and the commutator identity
\begin{equation}\label{}
(\c\,\dot{X})_{ab}=(\c\,X)^{\cdot}_{ab}+
  \frac{1}{3}(\c\,X)\,\Theta\;.
\end{equation}
These equations generalize the tensor perturbation equations for an
imperfect fluid that were derived in \cite{Challinor}. Once the form
of the anisotropic pressure has been determined in
Eq.~(\ref{eq2ordSigma})-(\ref{eq2ordElect}), the equations can be
solved to give the evolution of tensor perturbations. As already
noted in \cite{DBBE} the presence of a term that contains the shear
in  Eq.~(\ref{eq2ordElect}) makes this equation effectively third
order, so that it is not possible to write down a closed wave
equation for $E_{ab}$. If ${\pi}_{ab}=0$, it is easy to show that
for consistency, the solution for this field must also satisfy a
wave equation because the shear is a solution of a wave equation and
Eq.~(\ref{eqsigma}) holds. This will also be the case here because
in our case ${\pi}_{ab}\propto \sigma_{ab}$ and so the anisotropic
pressure will also satisfy a wave equation.

Following standard harmonic analysis, Eq.~(\ref{eq2ordSigma}) and
Eq.~(\ref{eq2ordMag}) may be reduced to ordinary differential
equations. It is standard~\cite{BDE} to use trace-free symmetric
tensor eigenfunctions of the spatial  the Laplace-Beltrami operator
defined by:
\begin{eqnarray}\label{eq:harmonic}
\3nab^{2}Q_{ab} = -\frac{k^{2}}{a^{2}}Q_{ab}\;,
\end{eqnarray}
where $k=2\pi a/\lambda$ is the wavenumber and $\dot{Q}_{ab}=0$.
Developing $\sigma_{ab}$ and $H_{a b}$ in terms of the $Q_{ab}$,
Eq.~(\ref{eq2ordSigma}) and Eq.~(\ref{eq2ordMag}) reduce to
\begin{equation}\label{eq2ordSigmaHarm}
\ddot{\sigma}^{(k)} + \frac{5}{3}\, \Theta \, \dot{\sigma}^{(k)} +
\left(\frac{1}{9}\, {\Theta }^2 + \frac{1}{6}{\mu} -\frac{3}{2}{p}
+\frac{k^{2}}{a^{2}}  \right)\, \sigma^{(k)}_ {ab}
=\dot{{\pi}}^{(k)}+ \frac{2}{3}\, \Theta \, {\pi}^{(k)}\;,
\end{equation}
\begin{equation}\label{eq2ordMagHarm}
\ddot{H}^{(k)} + \frac{7}{3}\, \Theta\, \dot{H}^{(k)} +
\frac{2}{3}\, \left({\Theta }^2 - 3\,{p}
+\frac{k^{2}}{a^{2}}\right)\, H^{(k)}= (\c\,
\dot{{\pi}})^{(k)}+\frac{2 }{3}\, \Theta\, (\c\, \ {\pi})^{(k)}\;,
\end{equation}
and Eq.~(\ref{eqsigma}) reads
\begin{equation}\label{eq2ordElectHarmLWL}
E^{(k)}=- \dot{\sigma}^{(k)} - \frac{2}{3}\,\Theta \,\sigma ^{(k)} +
\frac{1}{2} {\pi}^{(k)}\,.
\end{equation}

\section{General Equations for fourth order gravity}

The classical action for a fourth order theory of gravity is
given by
\begin{equation}\label{azione generale high order}
\mathcal{A} = \int d^4 x \sqrt{-g} \left[ \Lambda + c_{0} R + c_{1}
R^{2} + c_{2} R_{\mu \nu} R^{\mu \nu} + {\cal L}_{m} \right]\;,
\end{equation}
where we have used the Gauss Bonnet theorem \cite{GB}  and
$\mathcal{L}_m$ represents the matter contribution. In situations
where the metric has a high degree of symmetry, this action can be
further simplified. In particular, in the homogeneous and isotropic
case the action for a general fourth order theory of gravity takes the form
\begin{equation}\label{eq:actionFR}
\mathcal{A}=\int d x^{4}\sqrt{-g}\left[f(R)+\mathcal{L}_m \right]\;,
\end{equation}
where $\mathcal{L}_m$ represents the matter contribution. Such
modifications to the linear Einstein-Hilbert action can typically
arise in effective actions derived from higher-dimensional theories
of gravity~\cite{current4}. Varying the action with respect to the
metric give the gravitational field equations:
\begin{eqnarray}
f' R_{ab} - \sfrac{1}{2}g_{ab}f &=& \left( g^{c}{}_{a} g^{d}{}_{b}
-g_{ab}g^{cd} \right) S_{cd} + T_{ab}^{{\it m}}, \label{eq:einstFR}
\end{eqnarray}
where $f=f(R)$, $f'=f'(R)\equiv \partial f(R)/\partial R$,
$\displaystyle{T^{m}_{\mu\nu}=\frac{2}{\sqrt{-g}}\frac{\delta
(\sqrt{-g}\mathcal{L}_{m})}{\delta g_{\mu\nu}}}$ represents the
stress energy tensor of standard matter and
$S_{ab}=\nab_{a}\nab_{b}f'(R)$. The trace of Eq.~(\ref{eq:einstFR})
gives:
\begin{eqnarray}
f' R - 2f &=& -3S + T^{{\it m}}, \label{eq:einstTraceFR}
\end{eqnarray}
where $S=g^{ab}S_{ab}$. The various components of $S_{ab}$ can be
decomposed as
\begin{eqnarray}
S_{ab} &=& f''\left[ \tilde{\nab}_{a}\tilde{\nab}_{b}R -
\tilde{\nab}_{a}\dot{R}u_{b} -u_{a}u^{c}\nab_{c}(\tilde{\nab}_{b}R)
+\ddot{R} u_{a}u_{b} - \dot{R}\left(
\tilde{\nab}_{a}u_{b}-u_{a}\dot{u}_{b}\right) \right] \nonumber\\
\nonumber\\
&& +f'''\left[ \tilde{\nab}_{a}R \tilde{\nab}_{b}R - \dot{R}\left(
\tilde{\nab}_{b}R u_{a} +\tilde{\nab}_{a}Ru_{b} \right) +
\dot{R}^{2}u_{a}u_{b} \right],\nonumber\\
\nonumber\\
S &=& f''\left( \tilde{\nab}^{c}\tilde{\nab}_{c}R + \dot{u}^{c}
\tilde{\nab}_{c}R -\ddot{R}-\Theta \dot{R} \right) +f'''\left(
\tilde{\nab}^{c}R\tilde{\nab}_{c}R - \dot{R}^2 \right),
\end{eqnarray}
These equations reduce to the standard Einstein field equations when
$f(R)=R$. It is crucial for our purposes to be able to write
Eq.~(\ref{eq:einstFR}) in the form
\begin{equation}
\label{eq:einstFReff}
 G_{ab}=T^{tot}_{ab}= \tilde{T}_{ab}^{m}+T^{R}_{ab}\,,
 \end{equation}
where $\displaystyle{\tilde{T}_{ab}^{m}=\frac{ T_{ab}^{m}}{f'}}$ and
$T^{R}_{ab}$ is defined as
\begin{eqnarray}\label{eq:TenergymomentuEff}
T^{R}_{ab} &=& \frac{1}{f'}\left[ \frac{1}{2} \left( f-f' R \right)
g_{ab} + \left( g^{c}{}_{a} g^{d}{}_{b} -g_{ab}g^{cd} \right) S_{cd}
\right]. \label{eq:semt}
\end{eqnarray}

The RHS of Eq.~(\ref{eq:einstFReff}) represents two effective
``fluids":  the  {\em curvature ``fluid"} (associated with
$T^{R}_{ab}$) and  the {\em effective matter ``fluid"} (associated
with $\tilde{T}_{ab}^{m}$). This step is important because it allows
us to treat fourth order gravity as standard Einstein gravity in the
presence of two ``effective" fluids. This means that once the
effective thermodynamics of these fluids has been studied, we can
apply the covariant gauge invariant approach in the standard way.

The conservation properties of these effective fluids are given by
the Bianchi identities $T_{ab}^{tot\; ;b}$. When applied to the
total stress energy tensor, these identities reveal that if standard
matter is conserved, the total fluid is also conserved even though
the curvature fluid may in general possess off--diagonal terms
\cite{cdct:dynsys05,Taylor,eddington book}. In other words, no
matter how complicated the effective stress energy tensor
$T^{tot}_{ab}$ is, it will always be divergence free if
$T_{ab}^{m;b}=0$. When applied to the single effective tensors, the
Bianchi identities read
\begin{eqnarray}\label{Bianchi}
  &&\tilde{T}_{a b}^{m;b}=\frac{T_{ab}^{m;b}}{f'}-\frac{f''}{f'^{2}}\;T^{m}_{ab}\;R^{; b}\;,\\
  && T_{ab}^{R;b}=\frac{f''}{f'^{2}}\;\tilde{T}^{m}_{ab}\;R^{; b}\;,
\end{eqnarray}
with the last expression being a consequence of total
energy-momentum conservation. It follows that the individual
effective fluids are not conserved but exchange energy and momentum.

It is worth noting here that even if the energy-momentum tensor
associated with the effective matter source is not conserved,
standard matter still follows the usual conservation equations
$T_{ab}^{m\ ;b}=0$. It is also important to stress that the fluids
with $T^R_{ab}$ and $\tilde{T}^m_{ab}$ defined above are {\em
effective} and consequently can admit features that one would
normally consider unphysical for a standard matter field. This
means that all the thermodynamical quantities associated with the
curvature defined below should be considered {\em effective} and not
bounded by the usual constraints associated with matter fields. It
is important to understand that this does not compromise any of the
thermodynamical properties of standard matter represented by the
Lagrangian ${\cal L}_{m}$.

In the matter frame $u_a^m$ the various components of the total
energy momentum tensor, Eq.~(\ref{totEMTcomp}) can be re-written in
terms of the two effective fluids:
\begin{eqnarray}
{\mu} &=& \frac{\mu^m}{f'} + \mu^R, \\
\nonumber\\
{p} &=& \frac{p^m}{f'} + p^R, \\
\nonumber\\
{q}_{a} &=& \frac{q_{a}^m}{f'} + q_{a}^R, \\
\nonumber\\
{\pi}_{ab} &=& \frac{\pi_{ab}^m}{f'}+\pi_{ab}^R,
\end{eqnarray}
where we assume that standard matter is a perfect fluid, i.e.
$q_{a}^m =0$ and $\pi_{ab}^m =0$. The effective thermodynamical
quantities for the curvature fluid are
\begin{eqnarray}
&&\mu^{R}\,=\,\frac{1}{f'}\left[\frac{1}{2}(R f'-f)-\Theta
f''\dot{R}+f''\tilde{\nabla}^2{R}+f''\,\dot{u}_b\D{R}\right]\;,\\
&&p^{R}\,=\,\frac{1}{f'}\left[\frac{1}{2}(f-R
f')+f''\ddot{R}+f'''\dot{R}^2+\frac{2}{3}\Theta
f''\dot{R}-\frac{2}{3}f''\tilde{\nabla}^2{R}
-\frac{2}{3}f'''\D^{a}{R}\D_{a}{R}-\frac{1}{3}f''
\,\dot{u}_b\D{R}\right]\;,\\
&&q^{R}_a\,=\,-\frac{1}{f'}\left[f'''\dot{R}\D_{a}R+f''\D_{a}\dot{R}-\frac{1}{3}f''
\D_{a}R\right]\;,\\
&&\pi^{R}_{ab}\,=\,\frac{1}{f'}\left[f''\D_{\lgl
a}\D_{b\rgl}R+f'''\D_{\lgl a}{R}\D_{b\rgl}{R}-f'' \sigma_{a
b}\dot{R}\right]\,.\label{piR}
\end{eqnarray}
The twice contracted Bianchi Identities lead to evolution equations
for $\mu^{\,m}$, $\mu^{R}$, $q^{R}_a$:
\begin{eqnarray}
&&\dot{\mu}^m\,=\, - \,\Theta\,(\mu^m+{p^m})\;,\label{eq:cons1a}\\
&&\dot{\mu^R} + \3nab^{a}q^R_{a} + \,\Th\,(\mu^R+p^R) +
2\,(\udot^{a}q^R_{a}) +
(\sig^{a}\!^{b}\pi^R_{b}\!_{a})=\mu^{m}\frac{f''\,\dot{R}}{f'^{2}}\;,\label{eq:cons1b}\\
&&\dot{q}^R_{\lgl a\rgl} + \3nab_{a}p^R + \3nab^{b}\pi^R_{ab} +
\,{\textstyle\frac{4}{3}}\,\Th\,q^R_{a} + \sig_{a}\!^{b}\,q^R_{b} +
(\mu^R+p^R)\,\udot_{a} + \udot^{b}\,\pi^R_{ab} +
\eta_{a}^{bc}\,\om_{b}\,q^R_{c}=\mu^{m}\frac{f''\,\D_{a}{R}}{f'^{2}}\label{eq:cons1c}
\ , \end{eqnarray}
and a relation connecting the acceleration $\dot{u}_{a}$ to $\mu^m$
and $p^m$ follows from momentum conservation of standard matter:
\begin{equation}\label{eq:cons4}
\3nab^{a}{p^m} =  - (\mu^m+{p^m})\,\udot^{a}\,.
\end{equation}
Note that, as we have seen in the previous section the {\it
curvature} fluid and the effective {\it matter} exchange energy and
momentum. The decomposed interaction terms in Eq.~(\ref{eq:cons1b})
and Eq.~(\ref{eq:cons1c}) are given by
$\mu^{m}\frac{f''\,\D_{a}{R}}{f'^{2}}$ and
$\mu^{m}\frac{f''\,\dot{R}}{f'^{2}}$.

It is easy to see that the {\it curvature} fluid is in general an
imperfect fluid, i.e. has energy flux ($q_a$) and anisotropic
pressure ($\pi_{ab}$). Since we are only interested in linear tensor
perturbations, we need only be concerned with the tensor anisotropic
pressure, which is proportional to the shear, $\sigma_{ab}$. We now
present the second order evolution equations resulting from the
standard harmonic analysis of
Eq.s~(\ref{eq2ordSigmaHarm})-(\ref{eq2ordElectHarmLWL}) in the case
of f(R) theories of gravity:
\begin{eqnarray}\label{SigmaEqF}
\ddot{\sigma}^{(k)} + \left( \frac{5}{3}{\Theta}
+\dot{R}\frac{f''}{f'} \right)\dot{\sigma}^{(k)} + &&\left\{
\frac{1}{9}\, {{\Theta} }^2 + \frac{1}{f'}\left(\frac{1}{6}\mu^{m}
-\frac{3}{2}p^{m}\right) +\frac{k^2}{a^2} \right.\nonumber\\
&& \left.- \frac{1}{2}{\Theta}\dot{R}\frac{f''}{f'}
-\frac{5}{6}\frac{1}{f'}\left(f-f'R\right) -\dot{R}^{2} \left[
\frac{1}{2}\frac{f'''}{f'} + \left(\frac{f''}{f'}\right)^2 \right]
-\frac{1}{2}\ddot{R}\frac{f''}{f'} \right\} \sigma^{(k)}=0,
\end{eqnarray}
\begin{eqnarray}\label{HEqF}
\ddot{H}^{(k)} + \left( \frac{7}{3}{\Theta} +\dot{R}\frac{f''}{f'}
\right)\dot{H}^{(k)} +  &&\left\{ \frac{2}{3}\, {{\Theta} }^2
-\frac{2}{f'}p^{m} +\frac{k^2}{a^2} -\frac{1}{3}
{\Theta}\dot{R}\frac{f''}{f'} \right.\nonumber\\
&& \left. -\frac{1}{f'}\left(f-f'R\right) - \dot{R}^{2} \left[
\frac{f'''}{f'} + \left(\frac{f''}{f'}\right)^2 \right]
-\ddot{R}\frac{f''}{f'} \right\} H^{(k)}  = 0,
\end{eqnarray}
\begin{eqnarray}\label{EEqF}
&& {E}^{(k)} = -\dot{\sigma}^{(k)} - \left(\frac{2}{3}\Theta +
\frac{1}{2}\dot{R}\frac{f''}{f'} \right)\sigma^{(k)}.
\end{eqnarray}
For our purposes it will be particularly useful to consider these
equations in the so-called Long Wavelength Limit. In this limit the
wavenumber $k$ is considered to be so small that the wavelength
$\lambda=2\pi a/k$ associated with it is much larger than the Hubble
radius. Eq.~(\ref{eq:harmonic}) then implies that all the Laplacians
can be neglected and the spatial dependence of the perturbation
variables can be factored out.

\section{Tensor perturbations in $R^n$ gravity}

To proceed, we must now fix our theory of gravity, i.e. we must
choose the form of $f(R)$. We will consider a toy model
($R^n$-gravity) which is the simplest example of fourth order theory
of gravity but exhibits many of the properties of such theories. In
this theory $f(R)=\chi R^{n}$ and the action reads
\begin{equation}\label{71-curv1}
{\cal A}=\int d^4x \sqrt{-g} \left[\chi R^{n}+\mathcal{L}_{M}
\right]\;,
 \end{equation}
where $\chi$ a the coupling constant with suitable dimensions and
$\chi=1$ for $n=1$. If $R\neq 0$, the field equations for this theory
read
\begin{eqnarray}\label{equazioni di campo Rn}
G_{ab}=\chi^{-1}\frac{\tilde{T}_{ab}^{m}}{nR^{n-1}}+T^{R}_{ab}\;,
\end{eqnarray}
where
\begin{eqnarray}
\tilde{T}_{ab}^{m}&=&\chi^{-1}\frac{T_{ab}^{m}}{nR^{n-1}}\,,\\
T^{R}_{ab}&=&(n-1)\left\{-\frac{R}{2 n}g_{ab} +\left[\frac{R^{;c
d}}{R}+(n-2)\frac{R^{;c} R^{;d}}{R^{2}}\right](g_{c a}g_{d b}-g_{c
d}g_{ab})\right\}\,.
\end{eqnarray}
The FLRW dynamics of this model have been investigated in detail via
a dynamical systems approach in \cite{cdct:dynsys05}, where a
complete phase space analysis was performed. This work demonstrated
that for specific intervals of the parameter $n$ there exist a set
of initial conditions with non-zero measure for which the cosmic
histories include a transient decelerated phase (during which
large-scale structure can form) which evolves towards one with
accelerated expansion. These transient almost Friedmann models
existed for $0.28 \lesssim n \lesssim 1.35$ in the case of a dust
filled ($w=0$) universe and for $0.31 \lesssim n \lesssim 1.29$ in
the case of a radiation filled ($w=1/3$) universe. As we will
discuss in later sections, these allowed intervals of $n$ could be
reduced significantly with future observations of gravitational
waves. This model was also investigated as a possible explanation
for the observed flatness of the rotation curves of spiral galaxies
and the observed late times acceleration of the
universe~\cite{RotCurve}. The authors found a very good agreement
between this model and observational data when $n=3.5$. This is
however at odds with the results of~\cite{cdct:dynsys05}. Thus if
one requires a transient decelerated phase (during which large-scale
structure can form) and a solution to the dark matter and dark
energy problem, the $R^n$ model appears not to be viable. However
the aim of this paper is to show that the study of tensor
perturbations can in principle provide a strong constraint on the
theory of gravity independent of existing cosmological data sets and
consequently this work will provide a template for a more extensive
study of tensor perturbations of $f(R)$ cosmologies.

In what follows we begin by analyzing the evolution of tensor
perturbations in the absence of standard matter. We then consider
the case of dust/radiation dominated evolution. Although we will
give the full solutions, the discussion of the physics will be
restricted to the long wavelength limit.

\subsection{The vacuum case}\label{vacuum}

We start by considering tensor perturbations in the absence of
matter. This class of theories then admits the following exact
solution
\begin{eqnarray}
a(t) = a_{0}t^q, \qquad q=\frac{(1-n)(2n-1)}{n-2}, \qquad K=0.
\end{eqnarray}

\noindent The expansion parameter is given by
\begin{eqnarray}
\Theta(t) = \frac{3q}{t}.
\end{eqnarray}

For the purposes of this paper we restrict our attention to
expanding models. This requires $q>0$, which in turn restricts the
parameter $n$. In order to have an expanding background we require
$0<n<1/2$ and $1<n<2$ (we recover a static vacuum solution for
$n=1/2,~1$). We will only investigate models with values of $n$
which satisfy the second inequality (since we wish to investigate
models close to GR). The EOS of the total effective fluid in the
background is then \cite{revnostra}
\begin{eqnarray}
{w} &=& \frac{{p}}{{\mu}}
=-\frac{1}{3}\frac{(6n^2-7n-1)}{(2n-1)(n-1)}.
\end{eqnarray}
The EOS is singular and the pole's occur at $n=1/2$ and $n=1$.
Additionally, we have accelerated expansion, ($w<-1/3$) for
$n>(1+\sqrt{3})/2\approx 1.366$ and in the limit $n \to\infty$ we
have $w\to -1$. Substituting into Eq.'s
(\ref{SigmaEqF})-(\ref{EEqF}) we obtain
\begin{eqnarray}
\ddot{\sigma}^{(k)} + \frac{3(1-n)(4n-3)}{(n-2)t} \dot{\sigma}^{(k)}
+ \left[ \frac{n(4n-5)(n-1)(8n-7)}{(n-2)^2 t^2} + k^2 t^{-2q}
\right] \sigma^{(k)}=0,
\end{eqnarray}
\begin{eqnarray}
\ddot{H}^{(k)} + \frac{(1-n)(16n-11)}{(n-2)t} \dot{H}^{(k)} + \left[
\frac{2(6n^2 -8n+1)(n-1)(5n-4)}{(n-2)^2 t^2} + k^2 t^{-2q} \right]
H^{(k)}  = 0,
\end{eqnarray}
\begin{eqnarray}
&& {E}^{(k)} = -\dot{\sigma}^{(k)} +\frac{(n-1)(5n-4)}{(n-2)t}
\sigma^{(k)}.
\end{eqnarray}

In the long wavelength/super-horizon limit ($k=0$), the above
equations admit the following solutions
\begin{eqnarray}
\sigma^{(k)} &=& A_{1}t^{\frac{n(4n-5)}{n-2}} + A_{2}
t^{\frac{(8n-7)(n-1)}{n-2}}, \\
\nonumber\\
H^{(k)} &=& A_{3}t^{\frac{6n^2 -8n +1}{n-2}} + A_{4}
t^{\frac{2(5n-4)(n-1)}{n-2}}, \\
\nonumber\\
E^{(k)} &=& A_{1}(n-2)t^{\frac{2(n-1)(2n-1)}{n-2}} +
A_{2}\frac{(n-1)(5n-4)}{(n-2)} t^{\frac{8n^2 -16n+9}{n-2}}.
\end{eqnarray}

However, the physical quantity of interest is the dimensionless
expansion normalized shear $\Sigma=\sigma/H$.
\begin{eqnarray}\label{VacShear}
\Sigma^{(k)} &=& \tilde{\Sigma}_{1}t^{\frac{4n^2-4n-2}{n-2}}
+\tilde{\Sigma}_{2} t^{\frac{(2n-1)(4n-5)}{n-2}}.
\end{eqnarray}
\begin{figure}[t!]
\hspace{0.4cm}\includegraphics[width=8.5cm,angle=270]{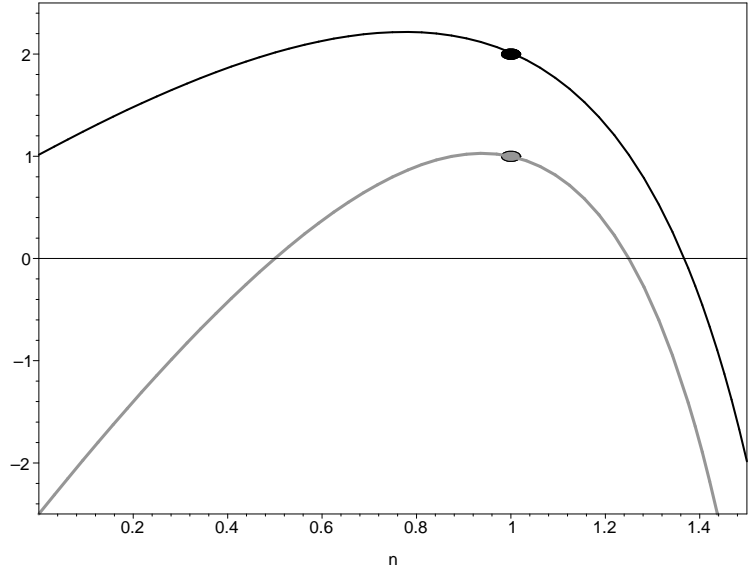}
 \caption{The exponents of each mode of the solution
Eq.~\ref{VacShear} against $n$ in the vacuum case. The black (grey)
line represents the growing (decaying) mode. The points represent
the value of the exponents in the case of GR ($n=1$). As $n$ is
increased the background expansion rate increase, resulting in a
weaker growth rate for tensor perturbations. When the critical value
of $n\approx 1.366$ is reached no growing modes can be supported in
this model.}\label{fig0}
\end{figure}
In Fig.~\ref{fig0} we have plotted the exponents of each mode of the
solutions given above as a function of $n$ in order to better see
how the large scale behavior varies. The black (grey) line
represents the growing (decaying) mode and the points represent the
value of the exponents in the case of GR ($n=1$). In the GR limit we
recover a static vacuum model in the background and
$\tilde{\Sigma}_{i}$ always grows indicating that this model is
unstable with respect to tensor perturbations. In the case of larger
values of $n$, the $\tilde{\Sigma}_{1}$ mode grows (decays) for
$n\lessapprox 1.366$ ($n \gtrapprox 1.366$) and the
$\tilde{\Sigma}_{2}$ mode grows (decays) for $n<1.25$ ($n>1.25$).
This is consistent with the background dynamics in that all
perturbation modes are decaying when we have accelerated expansion
(${w}<-1/3$) in the background.

For the sake of completeness we present the results of the general
case ($k\neq0$). The solutions are given in terms of Bessel
functions of the first and second kind ($J$ and $Y$ respectively).
\begin{eqnarray}
\sigma^{(k)} &=& t^{\frac{(2n-1)(6n-7)}{2(n-2)}} \left[ A_{1}
J\left(s,\frac{kt^r}{r}\right) + A_{2} Y
\left(s,\frac{kt^r}{r}\right) \right], \\
\nonumber\\
H^{(k)} &=& t^{\frac{(2n-1)(8n-9)}{2(n-2)}} \left[ A_{3}
J\left(s,\frac{kt^r}{r}\right) + A_{4} Y
\left(s,\frac{kt^r}{r}\right) \right], \\
\nonumber\\
E^{(k)} &=&  A_{1} t^{\frac{12n^2-22n+11}{2(n-2)}}
\left[\frac{n^2+2n-5}{(2-n)} J\left(s,\frac{kt^r}{r}\right) + kt^r
J\left(s+1,\frac{kt^r}{r}\right) \right] \\
&& +  A_{2} t^{\frac{12n^2-22n+11}{2(n-2)}}
\left[\frac{n^2+2n-5}{(2-n)} Y\left(s,\frac{kt^r}{r}\right) + kt^r
Y\left(s+1,\frac{kt^r}{r}\right) \right] ,
\end{eqnarray}
where we have introduced the following parameters
\begin{eqnarray}
\qquad r=\frac{2n^2-2n-1}{n-2}, \qquad s=-1+\frac{3(2n-3)}{2(n-2)r}.
\end{eqnarray}

The normalized shear $\Sigma$ is now of the form
\begin{eqnarray}
\Sigma^{(k)} &=& t^{\frac{3(4n^2 -6n+1)}{2(n-2)}} \left[
\tilde{\Sigma}_{1} J\left(s,\frac{kt^r}{r}\right) +
\tilde{\Sigma}_{2} Y \left(s,\frac{kt^r}{r}\right) \right].
\end{eqnarray}

\noindent where both the $\tilde{\Sigma}_{i}$ modes grow (decay) for
$n\lessapprox 1.290$ ($n \gtrapprox 1.290 $).

\subsection{The fluid case}\label{fluid}

We will now consider the case of tensor perturbations in the
presence of matter which is described by a perfect fluid with
barotropic EOS index, $w_m$. This class of theories then admits the
following exact solution
\begin{eqnarray}
a(t) = a_{0}t^{\frac{2n}{3(1+w_m)}}, \qquad K=0,
\end{eqnarray}

\noindent The expansion parameter is given by
\begin{eqnarray}
\Theta(t) = \frac{2n}{(1+w_m)t}.
\end{eqnarray}

As in the previous case we restrict our attention to expanding
models. Additionally, we are mainly interested in the case where the
perfect fluid describes dust ($w_m=0$) or radiation ($w_m=1/3$).
This is due to the fact that these cases are the most relevant when
considering GW detection via the CMB or direct detectors, e.g. LISA
and BBO. To insure an expanding model we now require $n>0$, provided
$w_m>-1$.

\begin{figure}[t!]
\hspace{0.4cm}\includegraphics[width=6.0cm,angle=270]{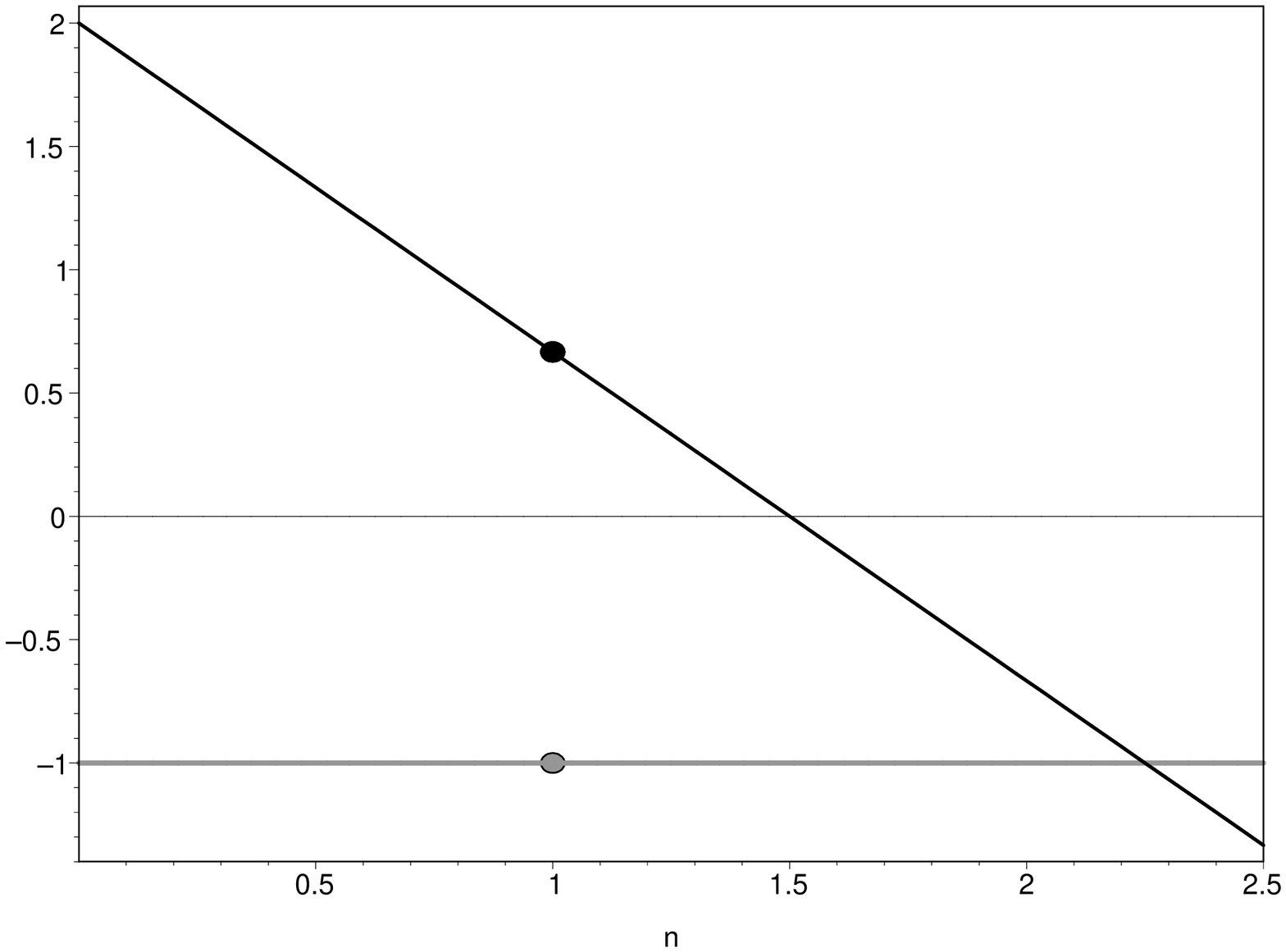}
\hspace{0.4cm}\includegraphics[width=6.0cm,angle=270]{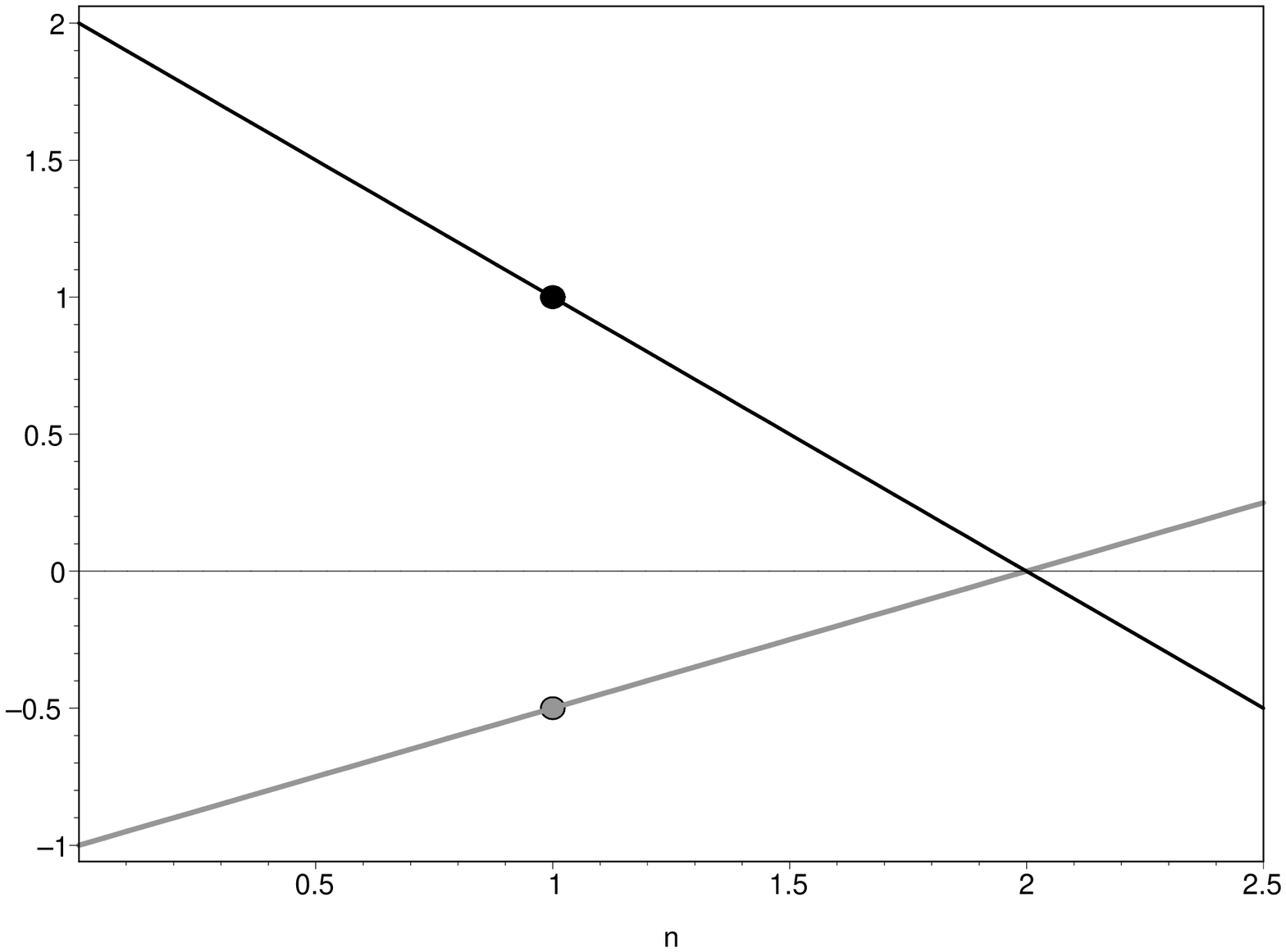}
\caption{The exponents of each mode of the solution for the
normalized shear against $n$ in the dust and radiation dominated
era's. (a) The left panel represents the exponents of the mode in
the dust dominated era. (b) The right panel represents the exponents
of the mode in the radiation dominated era. The black (grey) line
represents the growing (decaying) mode. The points represent the
value of the exponents in the case of GR ($n=1$).}\label{fig1}
\end{figure}

\subsubsection{The dust case}\label{dust}

We now investigate the evolution of tensor perturbations in the dust
dominated era. The scale-factor is given by
\begin{eqnarray}
a(t) = a_{0}t^{r},\qquad r=\frac{2n}{3}.
\end{eqnarray}
The EOS of the total effective fluid (dust and the effective
curvature fluid) is then
\begin{eqnarray}
w = -\frac{(n-1)}{n}.
\end{eqnarray}
The EOS is divergent for $n=0$ and we have accelerated expansion
($w<-1/3$) when $n>3/2$. In the limit $n \to \infty$ we have $w \to
-1$. Substituting into Eq.'s (\ref{SigmaEqF})-(\ref{EEqF}) we obtain
\begin{eqnarray}
\ddot{\sigma}^{(k)} + \frac{2(2n+3)}{3t} \dot{\sigma}^{(k)} + \left[
\frac{(8n-6)}{3 t^2} + k^2 t^{-2r} \right] \sigma^{(k)}=0,
\end{eqnarray}
\begin{eqnarray}
\ddot{H}^{(k)} + \frac{2(4n+3)}{3t} \dot{H}^{(k)} + \left[
\frac{2(2n^2 +5n-3)}{3 t^2} + k^2 t^{-2r} \right] H^{(k)} = 0,
\end{eqnarray}
\begin{eqnarray}
&& {E}^{(k)} = -\dot{\sigma}^{(k)} -\frac{(n+3)}{3t} \sigma^{(k)}.
\end{eqnarray}

\noindent In the long wavelength limit ($k=0$), the above equations
admit the following solutions
\begin{eqnarray}
\sigma^{(k)} &=& B_{1} t^{-2} +
B_{2} t^{\left(1-2r\right)}, \\
\nonumber\\
H^{(k)} &=& B_{3}t^{-\left(r+2\right)} + B_{4}
t^{\left(1-2n\right)}, \\
\nonumber\\
E^{(k)} &=& -B_{1}\frac{\left(9+n\right)}{3} t^{-3} - B_{2}
\frac{5n}{3} t^{-2r},
\end{eqnarray}
%

\noindent The normalized shear is given by
\begin{eqnarray}\label{DustShear}
\Sigma^{(k)} &=& \tilde{\Sigma}_{1}t^{-1} +\tilde{\Sigma}_{2}
t^{2\left(1-r\right)}.
\end{eqnarray}
The $\tilde{\Sigma}_{1}$ mode is the decaying mode solution and is
independent of the parameter $n$. This mode corresponds to the
standard decaying mode found in GR. The $\tilde{\Sigma}_{2}$ mode
grows (decays) for $n<3/2$ ($n>3/2$) and reduces to the GR growing
mode in the limit $n \to 1$. This is consistent with the background
dynamics in that all perturbation modes are decaying when we have
accelerated expansion (${w}<-1/3$) in the background.

In Fig.~\ref{fig1}(a) we have plotted the exponents of each mode of
the solutions given above as a function of $n$. The black (grey)
lines represents the growing (decaying) mode and the points
represent the value of the exponents in the case of GR ($n=1$). For
most of the values of $n$ the perturbations grow slower in
$R^n$-gravity than in GR. In fact only for $n<1$, does the
$\tilde{\Sigma}_{1}$ mode grow with a rate faster than the usual
$t^{2/3}$. In the case of GR, there is always a growing tensor
perturbation mode provided the background is not undergoing
accelerated expansion. In the case of $R^n$ gravity, tensor
perturbations grow at a slower rate, thus requiring a sufficiently
decelerated expansion in order to support a growing mode.

Again, for the sake of completeness, in the general case ($k\neq0$)
the solutions are given in terms of Bessel functions of the first
and second kind ($J$ and $Y$ respectively).
\begin{eqnarray}
\sigma^{(k)} &=& t^{-\left(2r+1\right)/2} \left\{ B_{1}
J\left[\frac{2r-3}{2\left(r-1\right)},
\frac{kt^{\left(1-r\right)}}{\left(r-1\right)}\right] + B_{2} Y
\left[\frac{2r-3}{2\left(r-1\right)},
\frac{kt^{\left(1-r\right)}}{\left(r-1\right)}\right] \right\}, \\
\nonumber\\
\nonumber\\
H^{(k)} &=&  t^{-\left(4r+1\right)/2} \left\{ B_{3}
J\left[\frac{2r-3}{2\left(r-1\right)},
\frac{kt^{\left(1-r\right)}}{\left(r-1\right)}\right] + B_{4} Y
\left[\frac{2r-3}{2\left(r-1\right)},
\frac{kt^{\left(1-r\right)}}{\left(r-1\right)}\right] \right\}, \\
\nonumber\\
\nonumber\\
E^{(k)} &=&  B_{1} t^{-\left(2r+3\right)/2}
\left\{\frac{2\left(2r-3\right)}{3}
J\left[\frac{2r-3}{2\left(r-1\right)},
\frac{kt^{\left(1-r\right)}}{\left(r-1\right)}\right] - kt^{-r}
J\left[\frac{1}{2\left(1-r\right)},
\frac{kt^{\left(1-r\right)}}{\left(r-1\right)}\right] \right\} \nonumber\\
\nonumber\\
&& +  B_{2}  t^{-\left(2r+3\right)/2}
\left\{\frac{2\left(2r-3\right)}{3}
Y\left[\frac{2r-3}{2\left(r-1\right)},
\frac{kt^{\left(1-r\right)}}{\left(r-1\right)}\right] - kt^{-r}
Y\left[\frac{1}{2\left(1-r\right)},
\frac{kt^{\left(1-r\right)}}{\left(r-1\right)}\right] \right\} ,
\end{eqnarray}

\subsubsection{The radiation case}\label{radiation}

Next, we study the evolution of tensor perturbations in the
radiation dominated era. The results of this section are especially
relevant if one wishes to constrain $f(R)$ models through their
impact on the B-mode correlation on the CMB. The scale-factor goes
as
\begin{eqnarray}
a(t) = a_{0}t^{r} \qquad r =\frac{n}{2}.
\end{eqnarray}
The EOS of the total effective fluid (radiation and the effective
curvature fluid) is then
\begin{eqnarray}
w = -\frac{\left(3n-4\right)}{3n}.
\end{eqnarray}
The EOS is divergent for $n=0$ and we have accelerated expansion
($w<-1/3$) when $n>2$. In the limit $n \to \infty$ we have $w \to
-1$. Substituting into Eq.'s (\ref{SigmaEqF})-(\ref{EEqF}) we obtain
\begin{eqnarray}
\ddot{\sigma}^{(k)} + \frac{n+4}{2t} \dot{\sigma}^{(k)} + \left[
\frac{\left(4-n\right)\left(n-1\right)}{2 t^2} + k^2 t^{-2r} \right]
\sigma^{(k)}=0,
\end{eqnarray}
\begin{eqnarray}
\ddot{H}^{(k)} + \frac{3n+4}{2t} \dot{H}^{(k)} - \left[
\frac{n\left( 3n-2\right)}{ t^2} + k^2 t^{-2r} \right] H^{(k)} = 0,
\end{eqnarray}
\begin{eqnarray}
&& {E}^{(k)} = -\dot{\sigma}^{(k)} -\frac{1}{t} \sigma^{(k)}.
\end{eqnarray}

\noindent In the long wavelength limit ($k=0$), the above equations
admit the following solutions
\begin{eqnarray}
\sigma^{(k)} &=& C_{1} t^{\left(1-2r\right)} +
C_{2} t^{\left(r-2\right)}, \\
\nonumber\\
H^{(k)} &=& C_{3}t^{\left(1-3r\right)} + C_{4}
t^{-2}, \\
\nonumber\\
E^{(k)} &=& C_{1}\left(2r-2\right) t^{-2r} + C_{2} \left(1-r\right)
t^{\left(r-3\right)},
\end{eqnarray}

\noindent The normalized shear is given by
\begin{eqnarray}\label{RadShear}
\Sigma^{(k)} &=& \tilde{\Sigma}_{1}t^{\left(2-2r\right)}
+\tilde{\Sigma}_{2} t^{\left(r-1\right)}.
\end{eqnarray}

\noindent The $\tilde{\Sigma}_{1}$ mode grows for $0<n<2$ and decays
for $n>2$. The $\tilde{\Sigma}_{2}$ mode decays for the range
$0<n<2$ and grows for $n>2$. In Fig.~\ref{fig1}(b) we have plotted
the exponents of each mode of the solutions given above as a
function of $n$. The black (grey) lines represents the growing
(decaying) mode and the points represent the value of the exponents
in the case of GR ($n=1$). For $0 < n <2$ the $\tilde{\Sigma}_{1}$
mode grows and the $\tilde{\Sigma}_{2}$ mode decays. In the range $n
> 2$ the modes change behavior in that the $\tilde{\Sigma}_{1}$ mode
decays and the $\tilde{\Sigma}_{2}$ mode grows. Again, for most of
the values of $n$ the perturbations grow slower in $R^n$-gravity
than in GR and only for $0 < n < 1$ and $n>4$, does the
$\tilde{\Sigma}_{i}$ modes grow with a rate faster than the usual
linear growth. The most interesting feature of the solutions in the
radiation dominated era is the possibility of growing modes even if
the universe is in a state of accelerated expansion ($n>4$). The
impact of these modes on the CMB, could allow one to constrain
deviations from GR. However, one should also analyze the evolution
of perturbations on small scales. This analysis is beyond the scope
of this paper and it is left to a future, more detailed
investigation.

In the general case ($k\neq0$) the solutions are given in terms of
Bessel functions of the first and second kind ($J$ and $Y$
respectively).
\begin{eqnarray}
\sigma^{(k)} &=& t^{-\left(r+1\right)/2} \left\{ C_{1}
J\left[\frac{3}{2},
\frac{kt^{\left(1-r\right)}}{\left(r-1\right)}\right] + C_{2} Y
\left[\frac{3}{2},
\frac{kt^{\left(1-r\right)}}{\left(r-1\right)}\right] \right\}, \\
\nonumber\\
\nonumber\\
H^{(k)} &=&  t^{-\left(3r+1\right)/2} \left\{ C_{3}
J\left[\frac{3}{2},
\frac{kt^{\left(1-r\right)}}{\left(r-1\right)}\right] + C_{4} Y
\left[\frac{3}{2},
\frac{kt^{\left(1-r\right)}}{\left(r-1\right)}\right] \right\}, \\
\nonumber\\
\nonumber\\
E^{(k)} &=&  C_{1} t^{-\left(r+3\right)/2} \left\{2\left(r-1\right)
J\left[\frac{3}{2},
\frac{kt^{\left(1-r\right)}}{\left(r-1\right)}\right] - kt^{-r}
J\left[\frac{5}{2},
\frac{kt^{\left(1-r\right)}}{\left(r-1\right)}\right] \right\} \nonumber\\
\nonumber\\
&& +  C_{2}  t^{-\left(r+3\right)/2} \left\{2\left(r-1\right)
Y\left[\frac{3}{2},
\frac{kt^{\left(1-r\right)}}{\left(r-1\right)}\right] - kt^{-r}
Y\left[\frac{5}{2},
\frac{kt^{\left(1-r\right)}}{\left(r-1\right)}\right]\right\} ,
\end{eqnarray}

\subsubsection{The generic large-scale case}\label{generic}

Finally, we study the evolution of large-scale ($k=0$) tensor
perturbations in the presence of a general barotropic fluid (thats
is we will not fix $w_m$ except to state that $w_m>-1$). To insure
an expanding model we now require $n>0$, provided $w_m>-1$. The
scale-factor goes as
\begin{eqnarray}
a(t) = a_{0}t^{r} \qquad r =\frac{2n}{3\left(1+w_m\right)}.
\end{eqnarray}
The EOS of the total effective fluid (radiation and the effective
curvature fluid) is then
\begin{eqnarray}
w = \frac{\left(w_{m}+1-n\right)}{n}.
\end{eqnarray}
The EOS is divergent for $n=0$ and we have accelerated expansion
($w<-1/3$) when $n>3(w_{m}+1)/2$. In the limit $n \to \infty$ we
have $w \to -1$. Substituting into Eq.'s
(\ref{SigmaEqF})-(\ref{EEqF}) we obtain
\begin{eqnarray}
\ddot{\sigma}^{(k)} +
\frac{10n+6\left(1+w_m\right)\left(1-n\right)}{3\left(1+w_m\right)t}
\dot{\sigma}^{(k)} + \frac{2\left(3+3w_m -4n\right)\left(n w_m - w_m
-1\right)}{3\left(1+w_m\right)^2 t^2} \sigma^{(k)}=0,
\end{eqnarray}
\begin{eqnarray}
\ddot{H}^{(k)} +
\frac{14n+6\left(1+w_m\right)\left(1-n\right)}{3\left(1+w_m\right)t}
\dot{H}^{(k)} + \frac{16n^2 -8n\left(w_m +1 \right)
+6\left(n-1\right)\left(w_m +1\right)\left(w_m +1 -
2n\right)}{3\left(1+w_m\right)^2 t^2} H^{(k)} = 0,
\end{eqnarray}
\begin{eqnarray}
&& {E}^{(k)} = -\dot{\sigma}^{(k)}
-\frac{4n+3\left(1+w_m\right)\left(1-n\right)}{3\left(1+w_m\right)t}
\sigma^{(k)}.
\end{eqnarray}

\noindent The solutions are then
\begin{eqnarray}
\sigma^{(k)} &=& D_{1} t^{\left(1-2r\right)} +
C_{2} t^{\left(2n-2-3r\right)}, \\
\nonumber\\
H^{(k)} &=& D_{3}t^{\left(1-3r\right)} + D_{4}
t^{2n-2-4r}, \\
\nonumber\\
E^{(k)} &=& D_{1}\left(n-2\right) t^{-2r} + C_{2} \left(n-1-r\right)
t^{\left(2n-3-3r\right)},
\end{eqnarray}

\noindent The normalized shear is given by
\begin{eqnarray}\label{RadShear}
\Sigma^{(k)} &=& \tilde{\Sigma}_{1}t^{\left(2-2r\right)}
+\tilde{\Sigma}_{2} t^{\left(2n-1-3r\right)}.
\end{eqnarray}

\begin{figure}[t!]
\hspace{0.4cm}\includegraphics[width=10.5cm,angle=0]{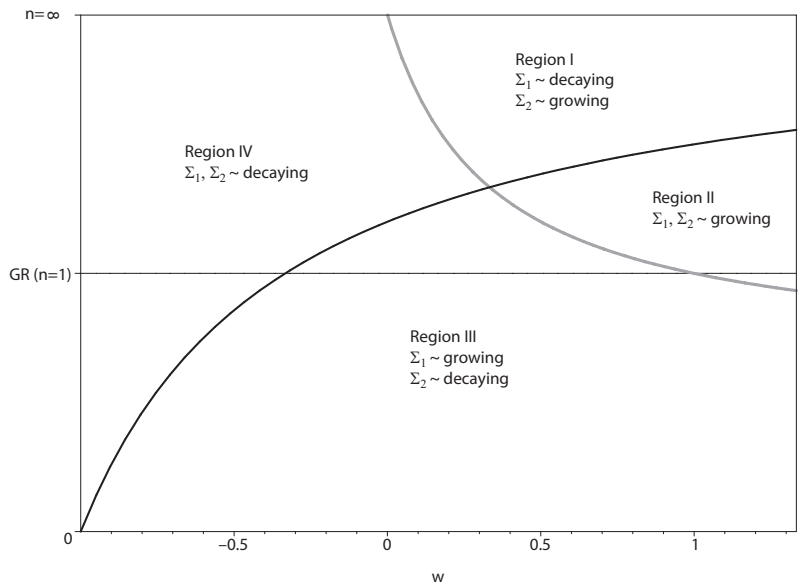} \caption{
The range of parameters for which the modes grow or decay as a
function of $n$ and $w_m$ The black thick line represents the change
from growth to decay for the $\tilde{\Sigma}_{1}$ mode. The grey
thick line represents the change from growth to decay for the
$\tilde{\Sigma}_{2}$ mode. The thin black line represents the GR
case ($n=1$). In region I, $\tilde{\Sigma}_{1}$ decays and
$\tilde{\Sigma}_{2}$ grows. In region II, both modes grow. In region
III, $\tilde{\Sigma}_{1}$ grows and $\tilde{\Sigma}_{2}$ decays.
Finally in region IV, both modes decay. }\label{fig2}
\end{figure}

The $\tilde{\Sigma}_{1}$ mode grows for $n<3(w_m +1)/2$ and decays
for $n>3(w_m +1)/2$. The $\tilde{\Sigma}_{2}$ mode decays for the
range $n<(w_m +1)/2w_m$ and grows for $n>(w_m +1)/2w_m$. In
Fig.~\ref{fig2} we have plotted the range of parameters for which
the modes grow or decay as a function of $n$ and $w_m$. This divides
the parameter space into four regions. In region I,
$\tilde{\Sigma}_{1}$ decays and $\tilde{\Sigma}_{2}$ grows. In
region II, both modes grow. In region III, $\tilde{\Sigma}_{1}$
grows and $\tilde{\Sigma}_{2}$ decays. Finally in region IV, both
modes decay. The most interesting feature of these solutions are
those of region IV. As mentioned earlier this particular model was
also investigated as a possible explanation for the observed
flatness of the rotation curves of spiral galaxies and the observed
late times acceleration of the universe~\cite{RotCurve}. The authors
found a good agreement between this model and observational
data when $n=3.5$ in the presence of dust ($w_m =0$). However, from
our analysis we have found that such a choice of parameters ensures
the absence of growing modes in the tensor perturbations. Therefore,
if we wish to use this model as an explanation for Dark Matter, we
can use gravitational wave detectors to severely constraint such
theories.

\section{Conclusions}

We have presented a mathematically well defined method of analyzing
the evolution tensor perturbations of FLRW backgrounds in fourth
order gravity, providing a general template for the study of linear
gravitational waves in this context. The analysis is based on two
important steps. Firstly, the recasting of the field equations for a
generic fourth order theory of gravity into a form which is
equivalent to GR, plus two effective fluids (the {\em curvature
``fluid"} and the {\em effective matter ``fluid"}). Secondly, using
1+3 covariant approach, it is possible to derive the general
equations describing the evolution of the cosmological perturbations
of these models for a FLRW background. In this paper we have only
dealt with the evolution of tensor perturbations,  the evolution of
scalar perturbations was been presented in \cite{Carloni:2007yv} and
the vector perturbations will be presented elsewhere \cite{K2}.
Providing that one has a clear picture in mind of the effective
nature of the fluids involved, the approach above has the advantage
of making the treatment of the perturbations physically clear and
mathematically rigorous.

Once the general perturbations equations were derived, we
specialized them to the case of the $R^{n}$-gravity model. Using
background solutions derived from an earlier dynamical systems
analysis \cite{cdct:dynsys05}, we found exact solutions to the
perturbation equations both in a vacuum and in the presence of matter
(dust and radiation). We presented both the large-scale limit and
full solutions, however, we restricted our discussions to the
large-scale results. In Section~\ref{vacuum} we studied the
evolution of tensor perturbations in vacuum. The background solution
proved to be unstable under tensor perturbations in the case of GR,
where the background represents a static vacuum solution. In
addition, for general values of $n$, the rate of growth of tensor
perturbations is weaker than the GR case.

In section~\ref{fluid} we studied the evolution of tensor
perturbations in the presence of matter. We first considered the
case of the dust dominated era. For most choices of $n$ ($n>1$) the
perturbations were found to grow at a slower rate in $R^n$-gravity
than in GR and no growing mode could be supported for $n>3/2$.

Next, we studied the evolution of tensor perturbation in the
radiation dominated era. Again, for most choices of $n$ ($n>1$) the
perturbations were found to grow at a slower rate in $R^n$-gravity
than in GR. However, it was found that their is always a growing
mode present except for the special case of $n=2$. This could have
important consequences on the tensor perturbation spectrum, e.g.
result in a tilt or running of the spectral index of the power
spectrum. In this way the connection between the spectrum of tensor
perturbations and the CMB polarization power spectrum offers an
interesting independent way of testing for alternative gravity on
cosmological scales.

Finally, we studied the evolution of tensor perturbation in the
presence of a generic fluid ($w_m >-1$) in the large scale limit
($k=0$). We found that there is a range of the parameters $w_m$ and
$n$ for which no growing modes are present ($n>3(w_m +1)/2$ and
$n<(w_m +1)/2w_m$). This corresponds the the choice of parameters as
required to solve the Dark Matter problem in the work of
~\cite{RotCurve}. Thus providing an alternate method of constraining
aforementioned theories of gravity.

As in the case of the results found for the evolution of the scalar
perturbations \cite{Carloni:2007yv}, the key question is how general
these results are in terms of the form of the fourth order
Lagrangian. Unfortunately this question is not easy to answer based
only on the analysis presented above. The key point to consider
would be the differences in the dynamics of the perturbations which,
as we have seen, are very pronounced but more difficult to used
because they depend largely on the features of the background. The
important point, however, is that these differences do not
necessarily imply a complete incompatibility with the data coming
from the CMB and other observational constraints. Much more work
will be needed before we can determine whether alternative gravity
provides a viable alternative to standard General Relativity.

\vfill {\noindent{\bf Acknowledgements:}\\
This work was supported by the National Research Foundation (South
Africa) and the {\it Ministrero deli Affari Esteri- DIG per la
Promozione e Cooperazione Culturale} (Italy) under the joint
Italy/South Africa science and technology agreement.

\appendix
\section{Covariant formalism versus Bardeen's formalism} \label{app1}

As we have seen the covariant approach is a very useful framework
for studying  perturbations in alternative theories of  gravity.
However, since most work on cosmological perturbations is usually
done using the Bardeen approach \cite{bardeen}, we will give here
a brief summary of how one can relate our quantities to the standard
Bardeen quantities. A detailed analysis of the connection between
these formalisms is given in \cite{BDE}. Here we limit ourselves to
give the main results for tensor perturbations.

In Bardeen's approach to perturbations of FLRW space-times, the
metric $g_{ab}$ is the fundamental object. If $\bar{g}_{ab}$ is the
background metric and $g_{ab}=\bar{g}_{ab} +\delta g_{ab}$ defines
the metric perturbations $\delta g_{ab}$ in these coordinates.

The perturbed metric can be written in the form
\begin{equation}
ds^2= a^2(\eta)\{-(1+2A)d\eta^2 -2 B_\alpha dx^\alpha d\eta
+[(1+2H_L) \gamma_{\alpha\beta}+2 H^T_{\alpha\beta}]dx^\alpha
dx^\beta\}\;, \label{eq:pertmetr}
\end{equation}
where $\eta$ is the conformal time, and the spatial coordinates are
left arbitrary. This spacetime can be foliated in 3-hypersurfaces
$\Sigma$ characterized by constant conformal time $\eta$ and metric
$\gamma_{ab}$.

The quantities $A$ and $B_\alpha$ are respectively the perturbation
in the lapse function (i.e. the ratio of the proper time distance
and the coordinate time one between two constant time hypersurfaces)
and in the shift vector (i.e. the rate of deviation of a constant
space coordinate line from the normal line to a constant time
hypersurface),  $H_L$ represents the amplitude of perturbation of a
unit spatial volume and $H^T_{\alpha\beta}$ is the amplitude of
anisotropic distortion of each constant time hypersurface
\cite{KodamaSasaki}.

The minimal set of perturbation variables is completed by defining
the fluctuations in the energy density: \ber  \mu=\bar{\mu}
+\delta\mu\;, & \delta\equiv\delta\mu/\bar{\mu}\;,
 \label{eq:dmu}
 \eer
and the fluid velocity: \ber u^a=\bar{u}^a +\delta u^a\;, & \delta
u^\alpha= \bar{u}^0 v^\alpha\;, & \delta u^0=-\bar{u}^0 A \;,
 \label{eq:du}
\eer
together with the energy flux $q_a$ and the anisotropic
pressure $\pi_{ab}$ which are GI by themselves.

These quantities are treated as 3-fields propagating on the
background 3-geometry. With suitable choice of boundary conditions
\cite{bi:stewart}, these quantities  can be uniquely (but
non-locally) decomposed into scalars, 3-vectors and 3-tensors: \ber
B_\alpha&=&B_{|\alpha} +B^S_\alpha\;,
 \label{eq:split1} \\
 H_{T\alpha\beta}&=&
\nab_{\alpha\beta}H_T +H^S_{T(\alpha|\beta)}
+H^{TT}_{T\alpha\beta}\;,
 \label{eq:split2}
 \eer
where the slash indicates covariant differentiation with respect to
the metric $\gamma_{ab}$ of $\Sigma$. In this way  $\nab_{ab}
f=f_{|\beta\alpha}-{\textstyle\frac{1}{3}} \nab^2f$ and
$\nab^2f=f^{|\gamma}{}_{|\gamma}$ is the Laplacian.  The superscript
$S$ on a vector means it is solenoidal ($B^{S|\alpha}_{\alpha}=0$),
and $TT$ tensors are transverse
($H^{TT\beta}_{T\alpha}{}_{|\beta}=0$) and trace-free.

On the base of \ref{eq:split1} and \ref{eq:split2}, it is standard
to define {\em scalar} perturbations as those quantities which are
3-scalars, or are derived from a scalar through linear operations
involving only the metric $\gamma_{ab}$  and its $|$ derivative.
Quantities derived from similar operations on solenoidal vectors and
on $TT$ tensors are dubbed  {\em vector} and {\em tensor}
perturbations. Scalar perturbations are relevant to matter clumping,
i.e. correspond to density perturbations, while vector and tensor
perturbations correspond to rotational perturbations and
gravitational waves.

Given the homogeneity and isotropy of the background, we can
separate each variable into its time and spatial dependence using
the method of harmonic decomposition. In the Bardeen approach the
standard harmonic decomposition is performed using the
eigenfunctions of the Laplace-Beltrami operator on 3-hypersurfaces
of constant curvature $\Sigma$ (i.e. on the homogeneous spatial
sections of FLRW universes). In particular these harmonics are
defined by
\begin{eqnarray}
&&\nabla^2 Y^{(k)}=-k^2 Y^{(k)}\;, \label{harmY1} \\
&&\nabla^2 Y^{(k)}_{\alpha}=-k^2Y^{(k)}_{\alpha}\;, \label{harmY2}\\
&& \nabla^2 Y^{(k)}_{\alpha\beta}=-k^2Y^{(k)}_{\alpha\beta}\;,
\label{harmY3}
\end{eqnarray}
where $Y^{(k)},Y^{(k)}_{\alpha},Y^{(k)}_{\alpha\beta}$ are the
scalar, vector and tensor harmonics of order $k$. In this way one
can decompose scalars, vectors and tensors as
\ber A&=&A(\eta)Y\\
B_\alpha&=& B\0(\eta) Y\0_\alpha +B\1(\eta) Y\1_\alpha\;,
 \label{eq:split1harm} \\
 H_{T\alpha\beta}&=&H\0_T(\eta)Y\0_{\alpha\beta}+H\1_T(\eta)
Y\1_{\alpha\beta} +H\2_T(\eta)Y\2_{\alpha\beta}\;.
\label{eq:split2harm} \eer
The key property of linear perturbation theory of FLRW space-times,
arising from the unicity of the splitting of (\ref{eq:split1}) and
(\ref{eq:split2}), is that in any vector and tensor equation the
scalar, vector and tensor parts on each side are separately equal,
i.e. the scalar, vector and tensor components of the equations
decouple.

All the quantities defined above can be decomposed in this way.
However, before proceeding, one should note that the quantities $A,
B_\alpha, H^L, H^T_{\alpha\beta}, \delta, v^\alpha$ change their
values under a change of correspondence between the perturbed
``world" and the unperturbed background, i.e., under a {\em gauge
transformation}. In order to have a gauge-invariant theory one has
to look for combinations of these quantities which are gauge
invariant. Bardeen constructed such GI variables to treat scalar and
vector perturbations \cite{bardeen}. The quantities which are
relevant to our analysis, $\pi_{\alpha\beta}$ and
$H_{T\alpha\beta}^{TT}$ (or the harmonically decomposed object
$H_{T}^{(2)}$) are already GI.

The  variables covariantly defined in the main text are, by
themselves, exact quantities (defined in any space-time) and are GI
by themselves, therefore, to first order, we can express them as
linear combinations of Bardeen's GI variables.  In \cite{BDE} these
expansions are given in full generality. Here we will limit
ourselves to a few examples, giving only the tensor contributions
and refer the reader to \cite{BDE} for details.

The tensor part of the shear, trace-free part of the 3-Ricci tensor,
the electric and magnetic parts of the Weyl tensor are given by
\begin{eqnarray}
\sigma_{\alpha\beta}&=& a {H^{(2)}_{T}}{}' Y_{\alpha\beta}^{(2)},
\label{eq:shear} \\
\nonumber\\
{}^{(3)}\mathcal{R}_{\alpha\beta}&=&
\left(k^2+2K\right)H^{(2)}_{T} Y\2_{\alpha\beta}, \label{eq:ricci}\\
\nonumber\\
E_{\alpha\beta} &=& -\frac{1}{2}\left[{H^{(2)}_{T}}{}''
-\left(k^2+2K\right)
H^{(2)}_{T}\right]Y\2_{\alpha\beta},   \label{eq:eweyl} \\
\nonumber\\
H_{\alpha\beta} &=& a^{-2} {H^{(2)}_{T}}{}'
Y\2_{(\alpha}{}^{\gamma|\delta}\eta_{\beta)0\gamma\delta}.
\label{eq:mweyl}
\end{eqnarray}
where the prime denotes derivative with respect to the conformal
time $\eta$. The relations above can be used to give an intrinsic
physical and geometrical meaning to Bardeen's variables, and also to
recover his equations. For example, combining our linearized
expression for the trace-free part of the 3-Ricci tensor
\begin{eqnarray}
{}^{(3)}\mathcal{R}_{\alpha\beta}&=&
-\frac{\Theta}{3}\left(\sigma_{\alpha\beta} + \omega_{\alpha\beta}
\right)+E_{\alpha\beta} +\frac{1}{2}\pi_{\alpha\beta},
\end{eqnarray}
with the above expressions (Eq.~\ref{eq:shear}-\ref{eq:eweyl}) gives
Bardeen's expression for the transverse and trace-free metric
perturbation evolution equation
\begin{eqnarray}
{H^{(2)}_{T}}{}'' +2\frac{a'}{a}{H^{(2)}_{T}}{}'+\left(k^2+2K\right)
H^{(2)}_{T}&=& \pi,
\end{eqnarray}
where $\pi$ is the harmonically decomposed anisotropic pressure.
Substituting for $\pi$ using Eq.~\ref{piR} and Eq.~\ref{eq:shear} we
find the general evolution equation for tensor perturbations in
fourth order gravity theories to be
\begin{eqnarray}
{H^{(2)}_{T}}{}'' +\left[2\frac{a'}{a}+\frac{ \partial^2 f}{\partial
R^2}\left(\frac{ \partial f}{\partial
R}\right)^{-1}R'\right]{H^{(2)}_{T}}{}'+\left(k^2+2K\right)
H^{(2)}_{T}&=& 0,
\end{eqnarray}
where primes denote differentiation with respect to conformal time
throughout this appendix.


\begin{thebibliography}{9999}


\bibitem{Grishchuk:1974ny}
 L.~P.~Grishchuk,
  Sov.\ Phys.\ JETP {\bf 40} (1975) 409
  [Zh.\ Eksp.\ Teor.\ Fiz.\  {\bf 67} (1974) 825];L.~P.~Grishchuk,
  Sov.\ Phys.\ Usp.\  {\bf 31} (1988) 940
  [Usp.\ Fiz.\ Nauk {\bf 156} (1988) 297].

\bibitem{Rubakov:1982df}
  V.~A.~Rubakov, M.~V.~Sazhin and A.~V.~Veryaskin,
  Phys.\ Lett.\ B {\bf 115} (1982) 189; A.A. Starobinsky,
  Pis'ma Astron. Zh.  {\bf 9}, 579 (1983) [Sov. Astron.
  Lett. 9, 302 (1983)]; L.~F.~Abbott and M.~B.~Wise,
  Astrophys.\ J.\  {\bf 282} (1984) L47;M.~J.~White,
  Phys.\ Rev.\ D {\bf 46} (1992) 4198
  [arXiv:hep-ph/9207239];M.~S.~Turner, M.~J.~White and J.~E.~Lidsey,
  Phys.\ Rev.\ D {\bf 48} (1993) 4613
  [arXiv:astro-ph/9306029].

  \bibitem{Hu:1997hv}
  W.~Hu and M.~J.~White,
  New Astron.\  {\bf 2} (1997) 323
  [arXiv:astro-ph/9706147].


\bibitem{revnostra}
 S.~Capozziello, S.~Carloni and A.~Troisi,
 ``Recent Research Developments in Astronomy \&  Astrophysics"-RSP/AA/21 (2003).
  [arXiv:astro-ph/0303041].

\bibitem{Odintsov}
  S.~Nojiri and S.~D.~Odintsov,
  Phys.\ Rev.\  D {\bf 68} (2003) 123512
  [arXiv:hep-th/0307288]


\bibitem{Carroll}
    S.~M.~Carroll, V.~Duvvuri, M.~Trodden and M.~S.~Turner,
  Phys.\ Rev.\  D {\bf 70}, 043528 (2004).
  [arXiv:astro-ph/0306438].

\bibitem{Capozziello:2005ku} S.~Capozziello, V.~F.~Cardone and A.~Troisi,
 Phys.\ Rev.\ D {\bf 71} (2005) 043503 [arXiv:astro-ph/0501426];

\bibitem{Capozziello:2006ph} S.~Capozziello, V.~F.~Cardone and A.~Troisi,
   Mon.\ Not.\ Roy.\ Astron.\ Soc.\ {\bf 375} (2007) 1423 [arXiv:astro-ph/0603522])

\bibitem{Capozziello:2006jj} S.~Capozziello, A.~Stabile and A.~Troisi,
Mod.\ Phys.\ Lett.\ A {\bf 21} (2006) 2291 [arXiv:gr-qc/0603071].

\bibitem{Capozziello:2006dj} S.~Capozziello, S.~Nojiri, S.~D.~Odintsov and A.~Troisi,
   Phys.\ Lett.\ B {\bf 639} (2006) 135 [arXiv:astro-ph/0604431]

\bibitem{cdct:dynsys05} S.~Carloni, P.~Dunsby, S.~Capozziello \& A.~Troisi
\cqg 22, 4839 (2005).

\bibitem{RotCurve}
  S.~Capozziello, V.~F.~Cardone and A.~Troisi,
  Mon.\ Not.\ Roy.\ Astron.\ Soc.\  {\bf 375} (2007) 1423
  [arXiv:astro-ph/0603522].



\bibitem{SanteGenDynSys}  S.~Carloni, A.~Troisi and P.~K.~S.~Dunsby,
  arXiv:0706.0452 [gr-qc] submitted to CQG.

\bibitem{shosho} M.~Abdelwahab, S.~Carloni and P.~K.~S.~Dunsby,
  arXiv:0706.1375 [gr-qc]. Submitted to CQG.

\bibitem{Carloni:2007yv}
  S.~Carloni, P.~K.~S.~Dunsby and A.~Troisi,
  arXiv:0707.0106 [gr-qc].

\bibitem{Will} C. M. Will,
, Living Rev. Relativity {\bf 9},  (2006),  3. URL (cited on
15/07/06): http://www.livingreviews.org/lrr-2006-3; {\em ibid}
in L.~J.~Dixon,
SLAC-R-538
{\it Prepared for 26th SLAC Summer Institute on Particle Physics:
Gravity -- From the Hubble Length to the Planck Length (SSI 98),
Stanford, California, 3-14 Aug 1998}

\bibitem{current1} R.~Bean, D.~Bernat, L.~Pogosian, A.~Silvestri and M.~Trodden,
  Phys.\ Rev.\  D {\bf 75}, 064020 (2007).
  [arXiv:astro-ph/0611321].

\bibitem{current2} Y.~S.~Song, W.~Hu and I.~Sawicki,
  Phys.\ Rev.\  D {\bf 75}, 044004 (2007).

\bibitem{current3}  B.~Li and J.~D.~Barrow,
  Phys.\ Rev.\  D {\bf 75}, 084010 (2007).
  [arXiv:gr-qc/0701111].

\bibitem{current4}
  K.~Uddin, J.~E.~Lidsey and R.~Tavakol,
  arXiv:0705.0232 [gr-qc].

\bibitem{bardeen}
J.~ M.~Bardeen,
 {\it Phys. Rev.} D, {\bf22}, 1982 (1980).

\bibitem{KodamaSasaki} H.~Kodama and M.~Sasaki,
  Prog.\ Theor.\ Phys.\ Suppl.\  {\bf 78} (1984) 1.

\bibitem{Bean2007} R.~Bean, D.~Bernat, L.~Pogosian,
A.~Silvestri and M.~Trodden,
  Phys.\ Rev.\  D {\bf 75} (2007) 064020
  [arXiv:astro-ph/0611321].

\bibitem{BDE}
M.~Bruni,  P.~K.~S.~Dunsby \& G.~F.~R.~Ellis,
Ap. J. {\bf 395} 34 (1992).

\bibitem{EB} G.~F.~R.~Ellis \& M.~Bruni
Phys Rev D {\bf 40} 1804 (1989).

\bibitem{EBH} G.~F.~R.~Ellis, M.~Bruni and J.~Hwang,
  Phys.\ Rev.\  D {\bf 42} (1990) 1035 (1990).

\bibitem{DBE}  P.~K.~S.~Dunsby, M.~Bruni and G.~F.~R.~Ellis,
  Astrophys.\ J.\  {\bf 395}, 54 (1992)

\bibitem{BED} M.~Bruni, G.~F.~R.~Ellis and P.~K.~S.~Dunsby,
 Class.\ Quant.\ Grav.\  {\bf 9}, 921 (1992).

\bibitem{DBBE}   P.~K.~S.~Dunsby, B.~A.~C.~Bassett and G.~F.~R.~Ellis,
  Class.\ Quant.\ Grav.\  {\bf 14}, 1215 (1997)
  [arXiv:gr-qc/9811092].

\bibitem{EllisCovariant} G.~F.~R.~Ellis \& H van Elst,
``Cosmological Models", Carg\`{e}se Lectures 1998, in Theoretical
and Observational Cosmology, Ed. M Lachièze-Rey, (Dordrecht: Kluwer,
1999), 1. [arXiv:gr-qc/9812046].

\bibitem{Challinor}
  A.~Challinor,
  Class.\ Quant.\ Grav.\  {\bf 17} (2000) 871
  [arXiv:astro-ph/9906474].

\bibitem{GB} I.~Chavel  {\em Riemannian Geometry: A Modern Introduction} (New York: Cambridge University Press, 1994).


\bibitem{Taylor} R.~Maarteens \& D.~R.~Taylor \grg {\bf 26} 599 (1994);

\bibitem{eddington book}A.~S.~ Eddington {\em The mathematical theory
of relativity} (Cambridge:  Cambridge Univ. Press 1952).

\bibitem{K2}   P.~ K.~ S.~ Dunsby, S.~Carloni  and K.~Ananda,  in preparation.

\bibitem{bi:stewart}
J.~M.~Stewart,  Perturbations of Friedmann\hs Robertson\hs Walker
cosmological models, {\it Class. Quantum Grav.} {\bf 7}, 1169
(1990).


\end{thebibliography}
\end{document}